\documentclass[twocolumn,floats,aps,floatfix,showpacs]{revtex4}
\usepackage{times} 
\usepackage{amssymb}
\usepackage{amsmath}
\usepackage{ulem}
\usepackage{color}
\usepackage{graphicx}     
\graphicspath{{.},{./Fig/}}
\newcommand{\mylab}[1]{\label{#1}}
\usepackage{ulem}
\newcounter{countuwe}

\newcounter{countcoaut}

\begin{document}
%
\title{Transport of free surface liquid films and drops by external ratchets and self-ratcheting mechanisms}
\author{Uwe Thiele}
\email{u.thiele@lboro.ac.uk}
\homepage{http://www.uwethiele.de}
\affiliation{Department of Mathematical Sciences, Loughborough University,
Loughborough, Leicestershire, LE11 3TU, UK}
\author{Karin John} 
\email{kjohn@spectro.ujf-grenoble.fr}
\affiliation{Laboratoire de Spectrom\'etrie Physique UMR 5588, CNRS
and Universit\'e Joseph Fourier - Grenoble I, BP 87 - 38402
Saint-Martin-d'H\`eres, France}
\begin{abstract}
  We discuss the usage of ratchet mechanisms to transport a continuous
  phase in several micro-fluidic settings. In particular, we study the
  transport of a dielectric liquid in a heterogeneous ratchet
  capacitor that is periodically switched on and off. The second
  system consists of drops on a solid substrate that are transported
  by different types of harmonic substrate vibrations. We argue that
  the latter can be seen as a self-ratcheting process and discuss
  analogies between the employed class of thin film equations and
  Fokker-Planck equations for transport of discrete objects in a
  'particle ratchet'.
\end{abstract}
%
%
\maketitle
%
%
%
%
\section{Introduction} \mylab{intro}
%
Most of the many studied ratchet mechanisms are discussed in
the context of directed transport and filtering of discrete objects
as, e.g., colloids or macromolecules \cite{HaMa09}.  Examples include
colloidal particles, that move in a directed manner through an array
of periodic asymmetric micropores when under the influence of an
oscillating external pressure \cite{MaMu03}. Such particles may as
well be driven by a dielectric potential of sawtooth shape that is
switched on and off periodically \cite{RSAP94}.  In many of those
'discrete' ratchets the carrier fluid (or solvent) does not or nearly
not move on average, i.e., they can not be employed to transport a
continuous phase. One exception is the motion of magnetic particles
in ferrofluids under the influence of an oscillating magnetic field
\cite{EMRJ03}. There the resulting net motion of the particles is
transmitted to the carrier liquid by a strong viscous coupling.

However, the basic concept of ratchet transport -- that a locally
asymmetric but globally homogeneous system may induce global transport
if it is kept out of equilibrium \cite{Curi1894} -- applies equally
well to pure continuous media. This has been employed recently to
transport a liquid or a solid phase in settings that do not show any
macroscopic gradient \cite{SISW03,QuAj06,BTS02}. The local asymmetry
can result, for instance, from a periodic but asymmetric external
potential. The periodic asymmetric variation needs to be on a small
length scale compared to the system size.

In a first example, drops that are either placed on an asymmetrically
structured substrate or between two such substrates move on average if
a transverse electric field is applied periodically \cite{BTS02}. A similar
transport may be achieved by vibrating the structured substrate in a tangential
direction \cite{BTS02}.
%
%
A second example is the triggering of a large scale mean flow in
{M}arangoni-{B\'e}nard convection over a solid substrate with
asymmetric grooves \cite{SISW03}. That is an interesting example (that
has not yet been studied in detail theoretically) as the static
ratchet profile of the substrate interacts with the time-periodic
motion of the convection rolls to produce the mean flow, i.e., the
time-periodic 'switching' is done by the system itself.  The strength
and direction of the mean flow depend on the thickness of the liquid
layer and the applied temperature gradient across the layer.
A third example are Leidenfrost drops that are placed on a hot surface
with a ratchet-like topography. This induces a directed motion of the
drops \cite{QuAj06,Link06}. The effect is not only observed for many
liquids but as well for small blocks of dry ice \cite{LLCQ10_pre}.

In contrast to the case of particle ratchets, not many models exist for the
ratchet-driven transport of a continuous phase.  Reference
\cite{Ajda00} considers a channel flow that is induced by locally
asymmetric periodic arrays of electrodes under an AC voltage
\cite{Ramo05}, and proposes an electro-osmotic model.

A different concept was analysed in relation to the first experimental
example: a two-layer film confined between two parallel plates can be
driven by periodically switching on and off an external potential that
is of ratchet shape in space. In the case that the potential is an
electrical one and the two plates form a (heterogeneous) capacitor one
may call the setup a flashing '(electro-)wettability ratchet'
\cite{JoTh07,JHT08}.  The concept of wettability may be understood in
a rather general way as it may include any effective
interactions between the liquid-liquid free interface and the solid
walls as long as one is able to apply them in a time-periodic, spatially
periodic (but locally asymmetric) manner.

Net transport of a continuous phase may not only be caused by an
external ratchet potential, but as well by a self-ratcheting effect:
An enlightening example are drops that are driven up an incline by
harmonic substrate vibrations at a finite angle to the substrate
normal \cite{BED07,BED09}. A simple ratchet-like mechanism has been
proposed: The vibration component that is orthogonal to the substrate
strongly modulates the hydrostatic pressure and thereby induces a
nonlinear response in the drop shape. That in turn determines the
strongly nonlinear drop mobility. As a result the drop reacts in an
asymmetric manner to the vibration component parallel to the
substrate. This symmetry breaking between back and forth motion leads
to the observed net motion of the drop.
A second experiment does not employ an oblique vibration but decouples
the normal and parallel vibrations entirely \cite{NKC09}. On a
horizontal substrate a net transport may be induced in either
direction depending on the phase-shift and amplitude ratio of the two
harmonic vibrations. Note, that in Ref.~\cite{NKC09} both vibrations
have the same frequency.
In both vibration experiments the phenomenon might be seen as a rocked
self-ratcheting \cite{HaMa09} of the drop as it is the drop itself
that introduces the local time-reflection asymmetry in the response to
the time-periodic driving of the sliding motion.

The decisive element in both presented systems -- the ratchet
capacitor and the vibrated drop -- is the interaction of the external
periodic forces with capillarity and wettability. With other words the
nonlinearity that is necessary for a net motion results from interface
effects.  One may call this class of ratchets, ``interfacial flow
driven ratchets''. As they are dominated by interface effects they
become more effective the smaller the involved scales are. This
implies that they are good candidates for micro- or even nano-fluidic
actuators.

In the present contribution we will restrict our attention to
one-layer films (i.e. liquid under gas) and formulate the thin film
model that applies to both driving types -- an external ratchet
potential and the vibration (Section~\ref{sec:model}).  Then we
present selected results for the external electrical ratchet potential
in Section~\ref{sec:ratchet}.
Section~\ref{sec:vibr:one} discusses drop transport by an oblique
vibration of the substrate (cf.~\cite{BED07}), whereas
Section~\ref{sec:vibr:two} describes results for the recently
investigated case where the drop moves as a reaction to a phase shift
between decoupled horizontal and vertical vibrations
\cite{NKC09}. Finally, we conclude and give an outlook in
Section~\ref{sec:conc}.

\section{Long-wave film profile evolution equation}
\mylab{sec:model}

%
\begin{figure}[tbh]
\begin{center}
{\large (a)\,}\includegraphics[width=0.6\hsize]{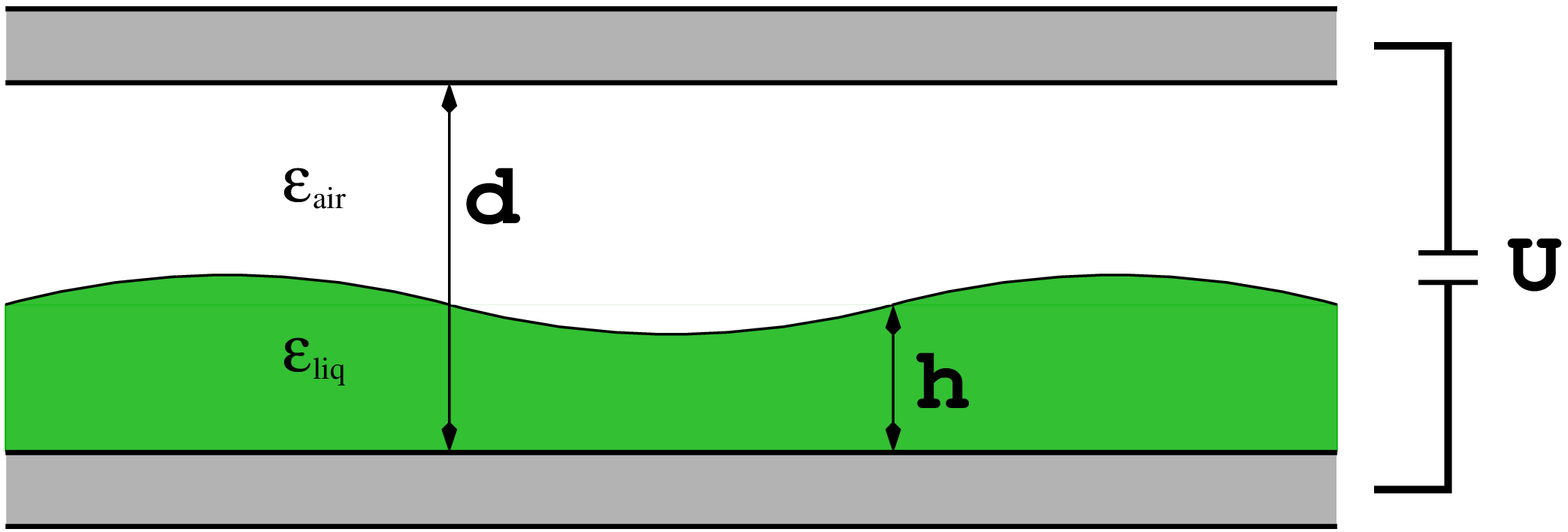}\\[2ex]
{\large (b)\,}\includegraphics[width=0.6\hsize]{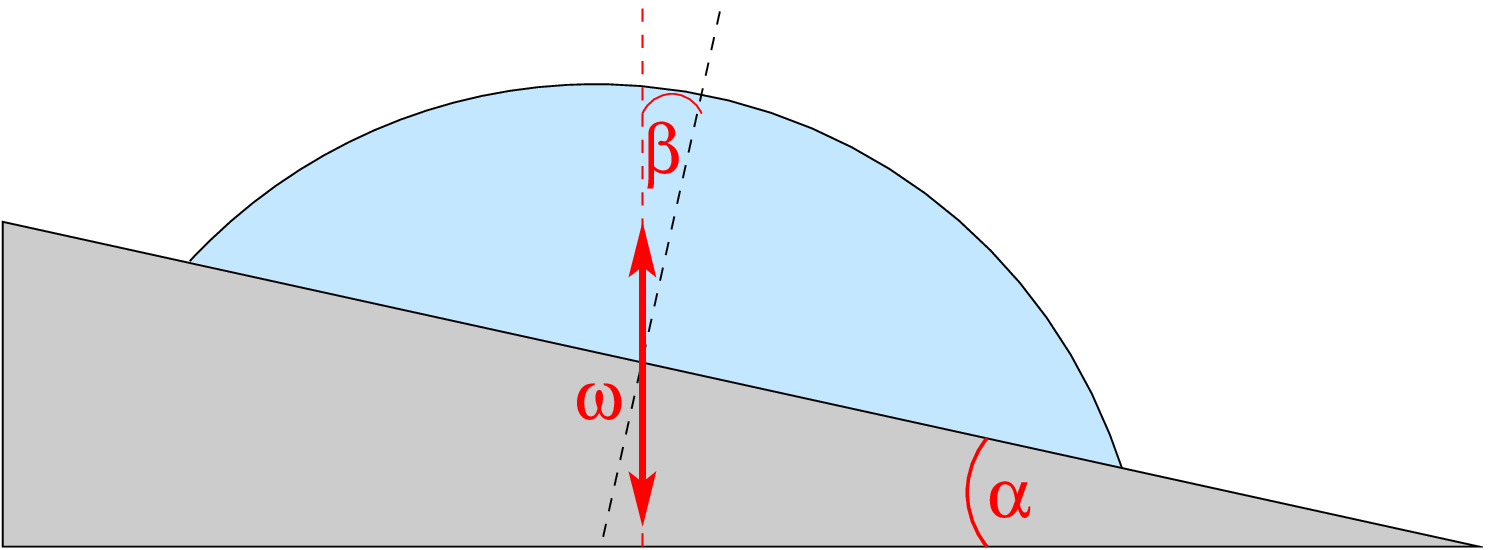}
\end{center}
\caption{(color online) (a) Sketch of a film of dielectric liquid of
  thickness $h(x)$ in a capacitor of gap width $d$ and voltage
  $U(x,t)$. The relative dielectric constants of the liquid and air
  are $\varepsilon_\mathrm{liq}$ and $\varepsilon_\mathrm{air}$, respectively.
  (b) Sketch of a drop on a vibrating inclined substrate. The
  frequency and angle of the vibration with the substrate normal are
  $\omega$ and $\beta$, respectively.  The inclination angle of the
  substrate is $\alpha$.\mylab{fig:sketch}}
\end{figure}

We restrict our attention to a two-dimensional system and consider
films/drops on lengthscales in the micrometer range.
%
%
The behaviour is then controlled by the interplay of the time-periodic
external force (and its spatial modulation), wettability and
capillarity. In the limit of small surface slopes (small contact
angles) it can be well described using an evolution equation for the
film thickness profile $h(x,t)$ that is derived from the momentum
transport equations with adequate boundary conditions 
employing long-wave approximation
\cite{ODB97,KaTh07}.  In dimensionless form we find
\begin{equation}
\partial_t\,h\,=\,-\partial_x\,\left\{Q(h)\partial_x\left[\partial_{xx}h + P(h,x,t)\right]
+ Q(h) F(t)\right\}.
\mylab{eq:film}
\end{equation}
As mass is conserved (no evaporation is considered), the time
derivative of the film thickness profile equals the divergence of a
flow. The flow results from a pressure gradient and a lateral force
and is proportional to a mobility $Q(h)$. The pressure
contains the Laplace or curvature pressure and a term $P(h,x,t)$ that
stands for all other pressure contributions as, e.g., disjoining
pressure (wettability), hydrostatic pressure, electrostatic pressure
and so on.  The pressure $P$ might depend on position and time. The
lateral driving force $F$ may include gravity (drop on incline),
thermal or wettability gradients and other. Depending on the
particular effect included it might as well depend on $h$.

\begin{figure}[ht]
\begin{center}
\includegraphics[width=0.8\hsize]{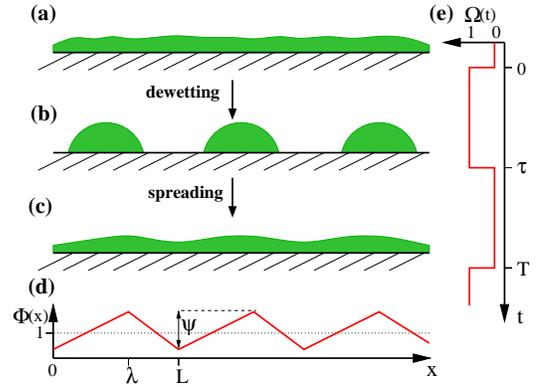}
\caption{(Color online) Panels (a) to (c) illustrate the working
principle of a fluidic ratchet based on a switchable wettability
that causes dewetting-spreading cycles.  (d) illustrates the
spatially asymmetric periodic interaction profile $\Phi(x)$
responsible for the wettability pattern and (e) indicates the time
dependence $\Omega(t)$ of the switching in relation to the
dewetting and spreading phases in (a) to (c).}
\mylab{fig:ratsche}
\end{center}
\end{figure}

We are here particularly interested in three cases:

(i) For a liquid film in an external electrical ratchet potential on a
horizontal substrate one has
 \begin{equation}
P(h,x,t)\,=\Pi_\mathrm{vdW}(h)\,+\,\Omega(t)\,\Phi(x)\,\Pi_{el}(h)
\mylab{eq:press}
\end{equation}
and $F=0$.  The disjoining pressure $\Pi_\mathrm{vdW}$ comprises the
effective interactions between the liquid-gas interface and the
substrate, i.e.\ the wettability properties
\cite{deGe85,Isra92,ODB97}.  We assume that a dielectric oil forms a
film in a capacitor of gap width $d$. The oil wets the lower plate and
does not wet the upper plate corresponding to
\begin{equation}
\Pi_\mathrm{vdW}=\left(\frac{A_l}{h^3}+\frac{A_u}{(1-h)^3}\right)
\end{equation} 
with the dimensionless Hamaker constants $A_l>0$ and $A_u<0$.
The dielectric film 
is subject to an electrical 'disjoining' pressure \cite{Lin01,MPBT05}
\begin{equation}
\Pi_{el} = \frac{(\varepsilon_r-1)}{[\varepsilon_r + 
(1-\varepsilon_r)h]^2}
\mylab{eq:djel}
\end{equation}
that varies in space as described by $\Phi(x)$ and is periodically
switched in time as $\Omega(t)$. In Eq.~(\ref{eq:djel}),
$\varepsilon_r$ denotes the ratio of the relative dielectric constants
of the liquid, $\varepsilon_\mathrm{liq}$, and the gas phase,
$\varepsilon_\mathrm{air}$.  The function $\Phi(x)$ models an electric field
that is periodic in $x$, but with an asymmetric profile, i.e., there
exists no reflection symmetry in $x$.  The employed spatial variation
$\Phi(x)$ and the temporal modulation $\Omega(t)$ are defined by the
sketches in Figs~\ref{fig:ratsche}(d) and \ref{fig:ratsche}(e),
respectively.

The introduced scales are
$3\gamma\eta/d\kappa^2_{el}$, $\sqrt{\gamma d/\kappa_{el}}$, and $d$
for time $t$, position $x$, and film thickness $h$, respectively.
Thereby, we have defined an electrostatic 'spreading coefficient'
$\kappa_{el}=\varepsilon_0\varepsilon_{\mathrm{liq}} U^2/2d^2$ where
$\varepsilon_0$ is the absolute dielectric constant and $U$ is the
applied voltage. $\gamma$ and $\eta$ denote surface tension and
dynamic viscosity of the liquid, respectively.

For simplicity we assume $A_l=-A_u=A>0$. The
dimensionless Hamaker constant $A$ is related to the dimensional one
by $A=A_\mathrm{dim}/6\pi d^3\kappa_{el}$. To characterise the ratchet
we
introduce the flashing ratio $\chi=\tau/(T-\tau)$ and the asymmetry
ratio $\phi=\lambda/(L-\lambda)$. The flashing frequency is
$\omega=2\pi/T$. The net transport along the substrate is measured by
the mean flow $\langle j \rangle$.  For more details see
\cite{JoTh07,JHT08}.

(ii) In the second example we look at drop transport by an oblique vibration
that determines 
both, $P$ and $F$. The pressure
\begin{equation}
P(h,x,t) =\Pi(h) - G h\left[1+a(t)\right]
\mylab{eq:press2}
\end{equation}
contains the disjoining pressure $\Pi(h)=-1/h^3+1/h^6$ and the
hydrostatic pressure where the time-dependence results
from the vibration component normal to the substrate. $\Pi$ contains
long-range destabilising 
and short-range stabilising 
van der Waals interactions \cite{Pism01}.
The lateral force 
\begin{equation}
F=G \left[\alpha+\beta b(t)\right]
\mylab{forceeq}
\end{equation}
contains a constant part (force down the incline) and a time-modulated
part (vibration component parallel to the substrate).  For a harmonic
oblique vibration, the substrate acceleration is $a(t)=b(t)=a_0\,\sin(\omega
t)$. Note, that the physical vibration angle is $O(\theta_e\beta)$
where $\theta_e$ is the mesoscopic equilibrium contact angle,
i.e., a scaled angle $\beta$ of order one corresponds to a small
physical angle.

In this case, the introduced scales are $3\gamma\eta/h_0\kappa^2$,
$\sqrt{\gamma h_0/\kappa}$, and $h_0=(B/|A|)^{1/3}$ for time $t$,
position $x$, and thickness $h$, respectively. $A<0$ and $B>0$ are the
Hamaker constants for the long- and the short-range part of the
disjoining pressure, respectively.  Furthermore, $\kappa=|A|/6\pi
h_0^3$, and $G=\rho g h_0/\kappa$, where $\rho$ denotes the density of
the liquid and $g$ the gravitational acceleration.  The
non-dimensional vibration period is $T=2\pi/\omega$.
The fixed drop volume
$V=L(\bar h-h_p)$ is determined by the domain size $L$, the mean film
thickness $\bar h$ and the dimensionless precursor film thickness $h_p=1$.  The
resulting transport along the substrate is measured after all
transients have decayed and the vibration-induced shape changes of the
drop are completely periodic in time. We quantify the transport by the
mean drop velocity $\langle v \rangle=\Delta x/T$ where $\Delta x$
denotes the distance the drop moves within one period $T$.

(iii) The final example is closely related to the second one. In the
experiments of Ref.~\cite{NKC09} the normal and lateral substrate
vibrations are mechanically decoupled and may have different
amplitudes, frequencies and phases, i.e., the normal vibration in
Eq.~(\ref{eq:press2}) is $a(t)=a_0\,\sin(\omega_a t +
\delta)$, and the lateral one in Eq.~(\ref{forceeq}) is $\beta
b(t)=\beta a_0\,\sin(\omega_b t )$. In the particular case of
\cite{NKC09}, a horizontal substrate is used, i.e., $\alpha=0$ and
$\omega_b=\omega_a=\omega$. The parameter $\beta$ takes the role of
the ratio of the vibration amplitudes in the directions parallel and
normal to the substrate.  To be consistent with the long-wave approach
taken, the physical amplitude ratio has to be small.  However, as in
(ii) $\beta$ is the scaled ratio and therefore of $O(1)$.

%
\section{Flow in a ratchet capacitor} \mylab{sec:ratchet}

Fig.~\ref{fig:ratsche} sketches an idealised electrical wettability
ratchet. The working principle is as follows. The flat free surface of
a film of a dielectric liquid that wets the lower wall of the
capacitor is stable when the electric field is switched off
(Fig.~\ref{fig:ratsche}\,(a)).
When switching on the spatially inhomogeneous electric field at $t=0$
(see spatial profile and time dependence in
Figs.~\ref{fig:ratsche}\,(d) and~\ref{fig:ratsche}\,(e), respectively)
the film dewets the substrate and decays into a set of drops (we call
this the 'on-phase'). This is due to a destabilisation of the
surface by the overall electric field {\it and} its gradients parallel
to the substrate (Fig.~\ref{fig:ratsche}\,(d)). The latter interfere
with the wavelength selection in the linear phase of the surface
instability. They do as well accelerate the coarsening process. The
different processes during the on-phase can be well appreciated
in Fig.~\ref{fig:ratsche2}.
The qualitative behaviour in the on-phase of the cycle resembles
dewetting of a liquid film on a substrate with a chemical wettability
pattern \cite{KaSh01,TBBB03}. If the on-phase is long enough
(as is the case in Fig.~\ref{fig:ratsche2}) all the liquid collects in
drops close to the positions of the maximal voltage
(Fig.~\ref{fig:ratsche}\,(b)). Then the field is switched off ('off-phase') at
$t=\tau$ (Fig.~\ref{fig:ratsche}\,(e)) and the drops spread
(Fig.~\ref{fig:ratsche}\,(c)), merge and become a homogeneous film
again (Fig.~\ref{fig:ratsche}\,(a)). The next cycle starts at
$t=T$. If the period of the cycle is not large enough, the surface may
not become entirely flat, a small modulation remains as indicated in
Fig.~\ref{fig:ratsche}(c).

When initially switching on the device there are several cycles
(typically $\sim 5-50$, the number varies with the employed
parameters) that show transient behaviour. However, the initial
transients die out and the evolution of the film profile during one
cycle is exactly time-periodic. Note, that there exist small parameter
ranges where this is not the case due to resonance phenomena and/or
multistability of various film states. However, this shall not concern
us here.

\begin{figure}[ht]
\begin{center}
\includegraphics[width=0.8\hsize]{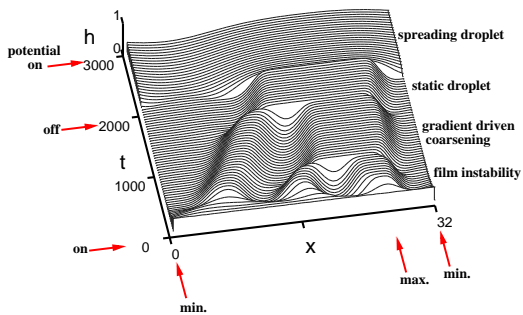}\\
\caption{Shown is a space-time plot of the typical evolution of the
  film thickness profile of a film of dielectric liquid in a
  heterogeneous capacitor (one spatial period shown) during one
  flashing period. For capacitor geometry and flashing cycle see
  Fig.~\ref{fig:ratsche}. Parameters are $\bar{h}=0.5$, $\Psi=0.5$,
  $L=32$, $\phi=5$, $T=5000$, $\chi=1$, $A=0.001$,
  $\varepsilon_r=2.5$. The starting time is well after initial
  transients have decayed.}  \mylab{fig:ratsche2}
\end{center}
\end{figure}

\begin{figure}[ht]
\begin{center}
\includegraphics[width=0.8\hsize]{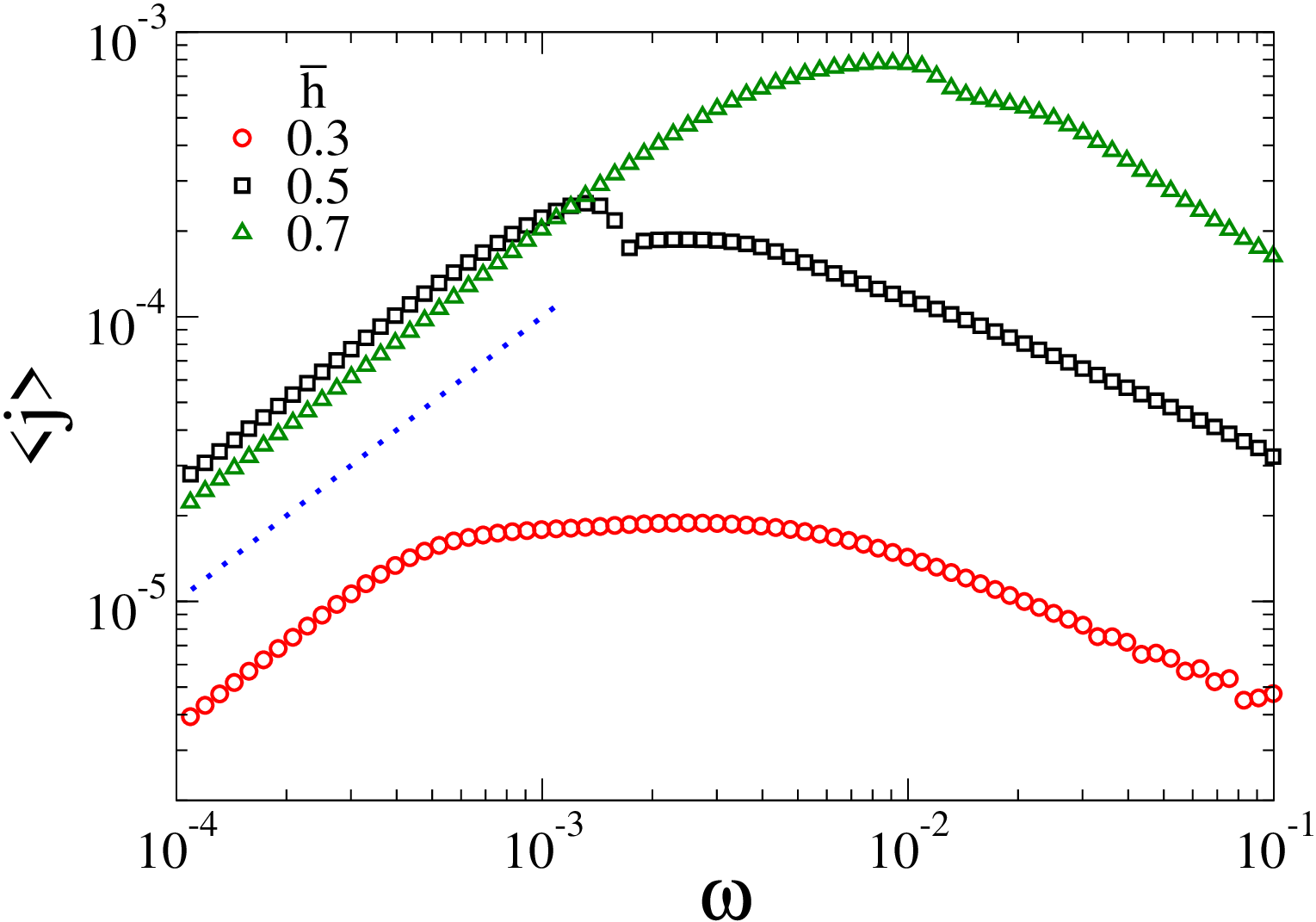}
\caption{(Color online) Shown is the net transport as measured by the
  mean flux $\langle j \rangle$ in its dependence on the flashing frequency
  $\omega$ for various film thicknesses as indicated in the legend.
  The remaining parameters are as in Fig.~\ref{fig:ratsche2}. The
  dotted straight line corresponds to a power law of exponent one.}
\mylab{fig:ratchet-flux}
\end{center}
\end{figure}

Although the evolution of the film profile is exactly time-periodic,
the process re-distributes the liquid within the film. This results in
a net transport that depends in a non-trivial way on the various
control parameters. As an example we give in
Fig.~\ref{fig:ratchet-flux} a log-log plot of the dependence of the
mean flux on the flashing frequency. For the used flashing ratio and
ratchet geometry the flux is always positive and approaches zero in
the zero frequency limit as well as in the high frequency limit.  For
large frequencies the fluid does not have enough time to dewet in the
on-phase or to spread in the off-phase.  At small
frequencies both processes reach the respective equilibrium well
before the next switching, i.e.\ most time is spend waiting.  As a
result the mean flow increases proportionally with frequency when
increasing the frequency from zero (cf.~Fig.~\ref{fig:ratchet-flux}).
The flux reaches a maximum at $\omega=10^{-3}\dots10^{-2}$ before it
decreases again. The decrease seems to follow a power law as well. It
is, however, not universal: the exponent increases with increasing
mean film thickness.  The decaying part shows non-monotonic step-like
behaviour that is more pronounced for smaller film thicknesses.  This
results from that fact that films of smaller thicknesses dewet in a
homogeneous electric field with a wavelength well below the spatial
period of the ratchet. The more wavelengths fit into a ratchet period
the more coarsening steps can take place in the on-phase if the
frequency is small enough. Each of the 'steps' in the curves of
Fig.~\ref{fig:ratchet-flux} is related to a coarsening step that does
not take place above the frequency at the step.
At smaller mean film thicknesses the maximum resembles a plateau with
a flux that does nearly not change between $\omega=10^{-3}$ and
$10^{-2}$.  The maximal flux increases with increasing mean film
thickness.  The particularly interesting non-monotonic behaviour close
to the flow maximum (here well visible for $\bar{h}=0.5$) results from
the interaction of the dynamics of the 'last' coarsening step and the
switching cycle.

Second, Fig.\,\ref{fig:chi} shows the dependence of the mean flux on
the flashing ratio. When it is low, i.e., the ratchet is most of the
time in the off-phase, the flux is very low.  This is because the
liquid remains nearly homogeneously spread out as it has no time to
assemble at the points of maximal voltage. By increasing the time of
the on-phase the flux increases in a non-trivial manner until it
reaches a maximum at $\chi\approx 1$. Depending on the film thickness
one may find almost a plateau, i.e. a range of flashing ratios with
nearly constant flux. For high flashing ratios the flux decreases
again, since the ratchet is most of the time in the on-phase. That
implies electrically formed drops have not enough time to spread out.
The decrease follows a power law with an exponent of about $-4/3$.

\begin{figure}[ht]
\begin{center}
\includegraphics[width=0.8\hsize]{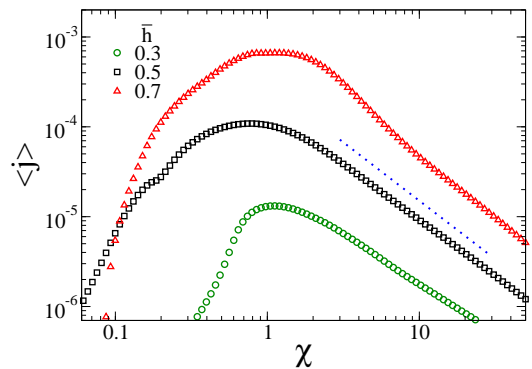}
\caption{(Color online) Shown is the net transport as measured by the
  mean flux $\langle j \rangle$ in its dependence on the flashing ratio
  $\chi$ for various film thicknesses as indicated in the legend.
  $T=500$ and the remaining parameters are as in
  Fig.~\ref{fig:ratsche2}. The dotted straight line corresponds to a
  power law of exponent -1.3.}  \mylab{fig:chi}
\end{center}
\end{figure}

Further calculations (not shown) indicate that the mean flux increases
monotonically with increasing spatial asymmetry ratio $\phi$ and the
amplitude $\Psi$ of the ratchet potential. Note that as expected the
net transport is zero at $\phi=1$, i.e., for a symmetric
potential. For further results for the presented one-layer geometry
see Refs.~\cite{JoTh07,JHT08}. Exchanging the air layer by a second
dielectric liquid one is able to transport the two liquids into
opposite direction. There, depending on the ratios of the
  viscosities and relative dielectric constants one may as well find a
  flux reversal.
This is further discussed in Ref.~\cite{JHT08}.
With this we close our present discussion of case (i), i.e., the study
of the transport of a dielectric liquid by a ratchet capacitor.  Our
results are typical for many externally imposed ratchets. The two cases
discussed next are both related to net transport through substrate
vibrations. Note that they can be seen as 'intrinsic' or
self-inflicted ratchets.

\section{Drop transport by oblique substrate vibration} \mylab{sec:vibr:one}

\begin{figure}[tbh]
\begin{center}
\includegraphics[width=0.7\hsize]{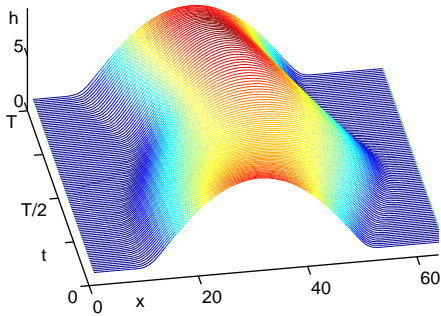}
\end{center}
\caption{(color online) Space-time plot illustrating the evolution of
  the profile of a drop on a obliquely vibrated substrate during one
  vibration period. Each cycle results in the net motion of the drop
  to the left. The starting time is well after initial transients have
  decayed. Note that only part of the domain $L$ is shown.  Parameters
  are $V=192$, $G=0.001$, $\beta=0.1$, $\alpha=0$, $T=400$
  ($\omega\approx0.017$), $L=128$, $a_0=10$.} \mylab{fig:evol1}
\end{figure}

As the first case of vibration-induced transport
 we discuss case (ii) introduced in Section~\ref{sec:model} -- a
drop transported by an oblique substrate vibration as sketched in
Fig.~\ref{fig:sketch}(b) and experimentally observed in Ref.~\cite{BED07}. 
The typical behaviour of a drop during one
vibration cycle as obtained in a time simulation of
Eq.~(\ref{eq:film}) is given in Fig.\,\ref{fig:evol1}.  One observes
that the drop undergoes changes of shape and moves back and forth.
Such simulations in time may be employed to analyse the net transport
over a wide frequency range. In addition one may use continuation
techniques \cite{DKK91,DKK91b,Thie02} to study the behaviour in the
low frequency limit, i.e., for a slowly vibrating substrate. In this
limit the intrinsic timescale of the drop dynamics $t_0$ is much
smaller than the vibration period $T$ and the drop moves in a
quasi-stationary manner. This means that drop shape and velocity at
each instant during the vibration cycle correspond to the ones of a
stationary moving drop at the corresponding constant force. They are
parametrized by $a(t)$. Averaging
stationary drop velocities $v(a(t))$ over one vibration period gives
the low frequency limit of $\langle v\rangle$.

\begin{figure}
\begin{center}
\includegraphics[width=0.7\hsize]{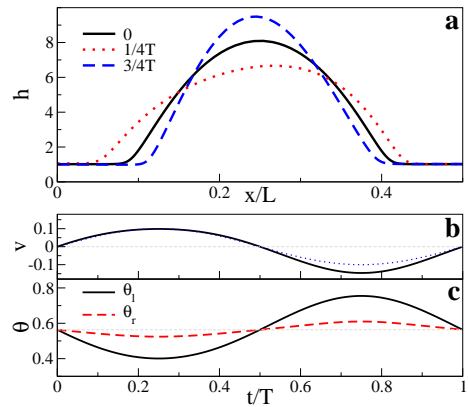}
\end{center}
\caption{(color online) Shown are for an obliquely vibrated horizontal
  substrate in the low frequency limit, (a) several drop shapes at
  times as given in the legends, (b) the momentaneous velocity, and
  (c) the left ($\theta_l$) and right ($\theta_r$) mesoscopic dynamic
  contact angle during one vibration cycle.  For the used $\beta=0.3$
  the drop moves with $\langle v \rangle \approx -0.01$, i.e., to the
  left. The blue dotted line in (b) indicates a harmonic variation of
  zero net flow.  Remaining parameters are as in Fig.~\ref{fig:evol1}.
  \mylab{fig:dropforms} }
\end{figure}

Fig.\,\ref{fig:dropforms} presents results for the low frequency
limit.  Panel (a) shows profiles of stationary moving drops at various
phases of the cycle on a horizontal substrate. Panels (b) and (c) give
drop velocity and the dynamic mesoscopic contact angles, respectively,
over one vibration cycle. The angles are shown for the left
($\theta_l$) and right ($\theta_r$) contact line. In the given setup,
temporal modulations of the right contact angle are larger than those
of the left one.  The deviation of the velocity of the sliding drop
from a harmonic modulation [dotted line in
Fig.~\ref{fig:dropforms}(b)] is a measure of the net motion of the
drop.

The overall behaviour of the drop over one vibration cycle is very
similar in Fig.~\ref{fig:evol1} (finite, but small frequency)
and~Fig.~\ref{fig:dropforms} (low frequency limit): during the first half
of the cycle ($t<T/2$) the drop is flattened and moves to the right.
Note that this happens when the substrate is accelerated upwards and
to the left.  In the second half of the cycle the drop becomes taller
and less wide while it slides to the left ($t>T/2$, substrate
acceleration is downward and to the right).  After one period the drop
has moved a small net distance to the left. For the parameters of
Fig.~\ref{fig:evol1} it takes about 100 vibration cycles to move the
drop by its own length. As detailed below in the conclusion, this does
well correspond to the available experimental results \cite{BED07}.

Based on the described findings one is able to understand the
mechanism that leads to the net motion of the drop: The component of
the oblique vibration that is normal to the substrate strongly
modulates the hydrostatic pressure and therefore provokes a nonlinear
response in the drop shape, i.e., when the substrate accellerates
upwards [downwards] it compresses [decompresses] the drop.  That in
turn determines the strongly nonlinear drop mobility $Q(h)$ in
Eq.~(\ref{eq:film}) and is therefore responsible for an asymmetric
response to the back and forth forcing that results from the parallel
vibration component. As a result of this nonlinear coupling of the
effects of normal and parallel vibration component one obtains an
anharmonic response of the drop to the harmonic but oblique vibration.

\begin{figure}
\begin{center}
\includegraphics[width=0.7\hsize]{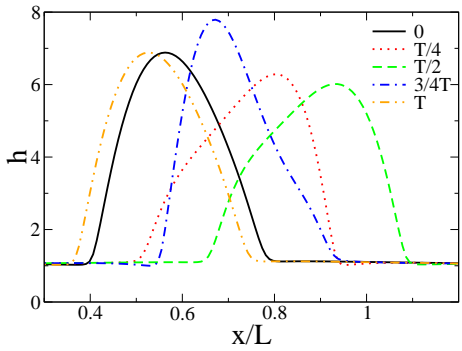}
\end{center}
\caption{(color online) Shown are several drop shapes during one
vibration cycle for an obliquely vibrated horizontal substrate at
$V=192$, $G=0.001$, $a_0=10$, $\alpha=0$, $T=400$, $L=128$ and a
relatively large vibration angle $\beta=1$. \mylab{fig:dropforms2} }
\end{figure}

The examples given in Figs.~\ref{fig:evol1} and~\ref{fig:dropforms}
where for a rather small vibration angle $\beta$. Therefore the change
in drop shape was mainly due to the modulation of the hydrostatic
pressure.  For larger $\beta$ the drop additionally changes its shape
in response to the parallel vibration component. An examples is given
in Fig.~\ref{fig:dropforms2}. The characteristic backwards shoulders
are known from strongly driven drops on homogeneous substrate
\cite{Thie01,Thie02}. For even larger $\beta$ the process may become
very complicated as the drop can during a single cycle undergo several
morphological changes. It can transform between a spherical cap-like
drop via a drop with a backwards shoulder to a finite film with a
capillary rim \cite{Thie01}. We do here, however, not investigate this regime
further.

\begin{figure}
\begin{center}
\includegraphics[width=0.7\hsize]{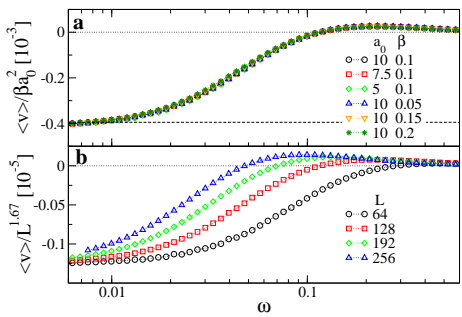}
\end{center}
\caption{Scaled mean velocity depending on the vibration 
frequency $\omega$ for drops on a horizontal substrate. 
Panel (a) gives the master curve for the scaled mean velocity $\langle v\rangle/\beta
  a_0^2$ for sets ($a_0$,
  $\beta$) as given in the legend. The horizontal dashed line
  indicates the result in the low frequency limit.  Remaining
  parameters are as in Fig.~\ref{fig:evol1}. 
  Panel (b) gives the scaled mean velocity $\langle v
  \rangle/V^{1.67}$ for drops of different volume as given in the
  legend.  Remaining parameters are $a_0=10.0$, $\beta=0.1$,
  $\alpha=0.0$, $G=0.001$.\mylab{fig:vT}}
\end{figure}

For smaller vibration angles $\beta$ morphological transitions do not
occur (at reasonable accellerations $a_0$). This allows for 'universal'
behaviour, i.e., one finds that scaling laws that are discovered
in the low frequency regime hold in part in the entire frequency range
studied. In particular, one finds in the low frequency regime a
scaling $\langle v \rangle \sim \beta a^2_0
L^{1.67}$. Fig.~\ref{fig:vT}(a) shows that for drops of identical
volume the scaling $\langle v \rangle \sim \beta a^2_0$ does hold very
well as it is possible to 'collaps' curves for a range of parameter
pairs $(a_0,\beta)$ on a single master curve. The scaling with volume
does, however, not hold [see Fig.~\ref{fig:vT}(b)].
Interestingly, one finds a flux reversal at high frequencies, i.e.,
above a critical frequency $\omega_c$ the drops move on average to the
right. Fig.~\ref{fig:vT}(b) further indicates that for larger drop volume the
reversal is more pronounced and occurs at lower frequencies $\omega_c$. We will
end this section with an investigation of the origin of the reversal.


\begin{figure}[hbt]
\begin{center}
\includegraphics[width=0.7\hsize]{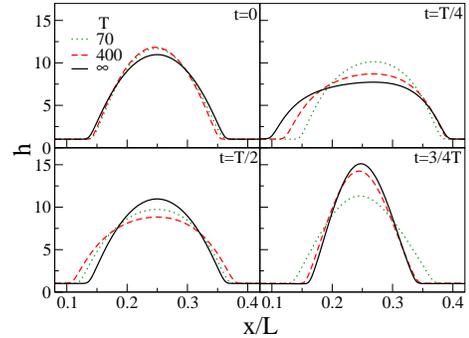}
\end{center}
\caption{Drop profiles at various times of the cycle as indicated in
upper right corner of each plot for various vibration periods $T$ as
indicated in the legend. $T=\infty$ indicates the low frequency
limit. Remaining parameters are $V=384$, $L=256$, $\alpha=0$,
$\beta=0.1$, $a_0=10$, $G=0.001$.
}
\mylab{fig:fluxreversalprofiles}
\end{figure}

\begin{figure}[hbt]
\begin{center}
\includegraphics[width=0.7\hsize]{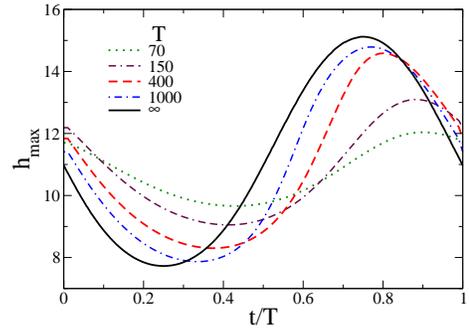}
\end{center}
\caption{Variation of the maximal drop height during one vibration period for
  different vibration periods as indicated in the legend. Remaining
  parameters are as in Fig.\,\ref{fig:fluxreversalprofiles}. }
\mylab{fig:fluxreversalhmax}
\end{figure}

To this end we show in Fig.\,\ref{fig:fluxreversalprofiles} a
superposition of drop profiles obtained at frequencies above and below
$\omega_c$ at selected times during the vibration cycle. The profiles
in the low frequency limit are per definition ideally in-phase with
the substrate vibration. At $T=400$
($\omega=0.016<\omega_c\approx0.05$, cf.~Fig.~\ref{fig:vT}(b)) the
profiles near the times of maximal accelleration (i.e., $t\approx
T/4$ and $t\approx3T/4$) strongly resemble the ones in the low
frequency limit, although the variation in their shape is already
smaller. For $T=1000$ they are nearly indistinguishable (not
shown). However, around $t=0$ and $T=T/2$ where the accelleration
changes sign the convergence to the low frequency limit is much slower
as there even small phase lags are very important
(cf.~Fig.~\ref{fig:fluxreversalhmax}).  The overall behaviour can be
well appreciated in Fig.~\ref{fig:fluxreversalhmax} where the
dependence of the maximal drop height on time is shown for one
vibration cycle for several different periods. 
With decreasing period we observe a continuous
increase of the phase shift w.r.t.\ the curve for the low frequency
limit. The curves of finite frequencies lag behind the one for zero
frequency and show a less pronounced variation of the drop heights.

At $T=70$ ($\omega=0.090>\omega_c$) the drop profiles in
Fig.\,\ref{fig:fluxreversalprofiles} vary much less over time than for
the lower frequencies and Fig.~\ref{fig:fluxreversalhmax} shows a
rather large phase shift (larger than $\pi/3$) as compared to the
other curves. 
When increasing the frequency, due to the phase shift and nonlinear
mobility eventually the average response to the backwards force
becomes smaller than the one to the forward force and the drop shows a
reversal of the net motion. The net motion to the right becomes
maximal at an optimal frequency. The influence of the phase lag still
exists at higher frequencies, however, as the changes in the profile
shape become smaller with increasing frequency, the mean velocity
decreases again and approaches zero in the limit of large
frequencies. This is as well the reason why the absolute values of the
velocities that can be reached in the flow reversal regime are much
smaller than the ones in the low frequency limit.  Note, finally, that
in the case of larger vibration angles the picture becomes more
complicated as the phase lag interacts with the morphological
changes. As a result the dependency of mean velocity on frequency
might have more than one maximum.

\section{Drop transport by decoupled horizontal and vertical substrate vibrations} 
\mylab{sec:vibr:two}

\begin{figure}
\begin{center}
\includegraphics[width=0.7\hsize]{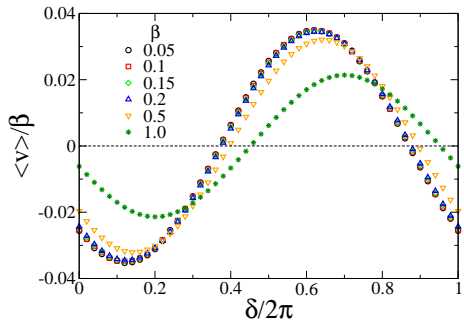}
\end{center}
\caption{(color online) Shown is the mean velocity as a function of
  the phase shift $\delta$ for various amplitude ratios $\beta$ as
  given in the legend.  Thereby, we use a relatively large period of
  $T=200$ and give the scaled velocity $\langle v\rangle/\beta$. For
  $\beta\lesssim0.2$ all curves 'collapse' onto a master curve.  The
  remaining parameters are $V=192$, $G=0.001$, $a_0=10$, $\alpha=0$,
  $L=128$, $a_0=10$.}  \mylab{fig:vel-on-phaseshift}
\end{figure}

As a second example of vibration-induced transport we discuss case
(iii) introduced in Section~\ref{sec:model} -- a drop transported on a
horizontal substrate by decoupled normal and parallel substrate
vibrations as experimentally observed in Ref.~\cite{NKC09}. We focus
on normal and parallel vibrations of identical frequency that have a
phase shift and possibly different amplitudes.  Following the
discussion of the underlyig mechanism in case (ii) it is no surprise
that a net motion is found with the present simple long-wave model as
well for case (iii).

Fig.~\ref{fig:vel-on-phaseshift} shows the dependence of the scaled
mean velocity on the phase shift for different amplitude ratios
$\beta$ at a moderately large vibration period of $T=200$
($\omega=0.031$). Very similar behaviour is found for other periods.
Interestingly, the dependencies of the scaled velocity $\langle
v\rangle/\beta$ for different $\beta$ fall on a single master curve
for $\beta\lesssim0.2$.  This implies that for $\beta\lesssim0.2$
particularly important values of the phase shift do not depend on the
amplitude ratio. In particular, we find that flow reversal occurs at
$\delta=3/4\pi$ and $\delta=7/4\pi$. The drop moves fastest to the
left [right] at $\delta=\pi/4$ [$\delta=5/4\pi$].  The master curve
shows the symmetry $(\delta\rightarrow
\delta+\pi,\langle v \rangle/\beta\rightarrow-\langle v\rangle/\beta)$.  Note that
the symmetry holds for all considered amplitude ratios and periods
(cf.~Fig.~\ref{fig:vel-on-phaseshift}), as can be expected from the
set-up of the problem.

\begin{figure}
\begin{center}
\includegraphics[width=0.7\hsize]{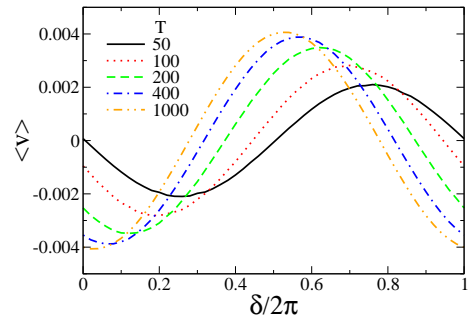}
\end{center}
\caption{(color online) Shown is the mean velocity as a function of
  the phase shift $\delta$ for various periods as given in the legend.
  The remaining parameters are $V=192$, $G=0.001$, $a_0=10$,
  $\beta=0.1$, $\alpha=0$, $L=128$, $a_0=10$.}
  \mylab{fig:vel-on-phaseshift2}
\end{figure}
However, the scaling found for small $\beta$ does not hold for larger
$\beta$ (see Fig.~\ref{fig:vel-on-phaseshift}).  For
$0.2\lesssim\beta\lesssim 1.0$ the phase shift that results in maximal
net motion and the maximal net velocity $\langle v \rangle$ both
increase with increasing $\beta$. For $\beta\gtrsim 1.0$ both decrease
again (not shown). This implies that the amplitude ratio that
maximises net transport is $\beta=1$. There the phase shift resulting
in maximal net transport to the left is slightly smaller than
$\delta=\pi/2$.
%

The change of the dependency of transport on phase
shift with changing period is shown in
Fig.~\ref{fig:vel-on-phaseshift2}. In accordance with results in case
(ii) one finds that the velocities become monotonically smaller (for
all phase shifts) with increasing frequencies (decreasing periods) as
the drops are less able to follow the substrate vibrations. The phase
shift where the maximal mean velocity occurs becomes smaller with
decreasing frequencies. In the limit of low frequency the fastest
transport to the left and right are found at $\delta=0$ and
$\delta=\pi$, respectively.

A comparison with the experiments in Ref.~\cite{NKC09} (in particular
their Fig.~3) shows that there the phase shift value where the maximal
mean velocity is found and the value of the maximal mean velocity both
increase with increasing ratio of the amplitudes of parallel and
normal vibration. However, as further explained in the conclusion, a direct
quantitative comparison is not possible. The experimental
results show roughly the symmetry
$(\delta\rightarrow\delta+\pi,\langle v\rangle\rightarrow-\langle v\rangle)$. Small deviations
are explained by the presence of defects on the substrate \cite{NKC09} 
(see, in particular, the inset of their Fig.~2).
A striking difference between the results presented here and Fig.~3(a)
of \cite{NKC09} is that in the latter the mean velocity depends in a
strongly non-harmonic way on the phase shift. In the simulations we
have not seen the experimentally found double peak structure around
the maximal velocity. This point needs further investigation.

\section{Conclusions}
\mylab{sec:conc}

We have explored the usage of ratchet mechanisms to transport a
continuous phase in micro-fluidic settings involving a free liquid-gas
interface and contact lines, i.e., under the influence of capillarity
and wettability. In particular, we have studied on the one hand the
transport of a dielectric liquid in a capacitor with an asymmetrically
spatially modulated electrical field that is periodically switched on
and off. In this case the ratchet-like potential is imposed
externally.
On the other hand we have investigated drops on a solid substrate that
show net motion under the influence of different types of harmonic
substrate vibrations. Analysing the underlying mechanism we have found
that the component of the harmonic vibration that is orthogonal to the
substrate induces a nonlinear (anharmonic) response in the drop shape.
The latter determines the strongly nonlinear drop mobility what
results in an asymmetric response of the drop to the vibration
component that is parallel to the substrate. The induced symmetry
breaking between forward and backward motion during the different
phases of the vibration results in the observed net motion of the
drop.
We have argued that the phenomenon might be seen
as a rocked self-ratcheting as the drop itself introduces the local
time-reflection asymmetry in the response to the time-periodic driving
of the sliding motion. We employ the term ``rocked'' as the net motion
is strongest in the low frequency limit as is typical for rocked
particle ratchets \cite{HaMa09}. Note as well that in the context of a
drop on an incline the component of the vibration parallel to the
substrate can be seen as a periodic rocking of the substrate.

For the vibrated drops our focus has been on two experimentally
realised cases \cite{BED07,BED09,NKC09}: an oblique substrate
vibration and decoupled normal and parallel substrate vibrations of
identical frequency but different amplitudes and/or phase shifts. As
in the latter case the use of different amplitudes without phase shift
corresponds to the case of the oblique vibration we have mainly
analysed the dependence of the drop motion on phase shift.

The decisive element in both presented systems -- the ratchet
capacitor and the self-ratcheting vibrated drop -- is the interaction
of the external periodic forces with capillarity and wettability,
i.e., the nonlinearity that is necessary for a net motion results from
interface effects.  One may call this class of ratchets, ``interfacial
flow driven ratchets''. As they are dominated by interface effects
they are very effective on small scales. This implies that they are
good candidates for micro- or even nano-fluidic actuators.

One needs to keep in mind that all models analysed in the present
contribution are obtained employing a long-wave or lubrication
approximation \cite{ODB97,KaTh07}. That means they are relatively
simple, but fully dynamical, highly nonlinear models based on a
minimal set of ingredients, i.e., Stokes flow in the lubrication
approximation, (electro-)capillarity, and wettability.  They allow to
study the net transport in the ratchet capacitor and the net motion of
the vibrated droplet over a wide range of parameters, to understand the
underlying mechanisms and to discuss transitions in the qualitative
system behaviour.

However, when comparing such lubrication models with particular
experiments one always has to keep in mind that the formal range of
applicability of the lubrication approximation is limited and does
often not overlap with the parameter ranges where experiments work
best. The latter are often performed with liquid-substrate combinations
that lead to equilibrium (and dynamic) contact angles that are not
small. In our case (ii) one uses as well large substrate inclinations
and large angles between vibration direction and substrate normal. 
For a direct quantitative comparison \textit{all} those angles have to be small.
Nevertheless, lubrication models are extremely successful in
explaining intriguing effects observed in a wide range of experiments
involving capillarity and wettability. A good example for this are
studies of morphological changes in sliding drops where some
experiments are done for liquids with static contact angles above
$\pi/4$ using substrate inclinations varying from zero to $\pi/2$
\cite{PFL01}. Lubrication theory well explains the effects (see, e.g.,
\cite{Snoe07} and the discussion in \cite{Bonn09}).

For the vibrated drop in Ref.~\cite{BED07} a liquid is used that has
an equilibrium contact angle of about $\pi/3$, the substrate
inclination is $\pi/4$, and the vibration is applied at an angle of
$\pi/4$ with the substrate normal, i.e., the experiment is well
outside the formal range of applicability of the lubrication
approximation. Therefore, a direct quantitative comparison of mean
velocities could only be done through some 'up-scaling' procedure that
lifts drop volume, contact angles and vibration angle from the
lubrication theory values to the experimental ones. This might be done
(and is sometimes done). However, in our opinion the process has
arbitrary elements, and we prefer not to employ it.

The theory may, however, be related to the experiment in a
semi-quantitative way by comparing, for instance, mean velocity in
terms of drop size or typical timescales. In Ref.~\cite{BED07} the
employed frequencies $f$ range from 25\,Hz to 120\,Hz. At an
acceleration of 15\,$g$ and $f=60$\,Hz one finds mean velocities of about
1mm/s, i.e., the drop moves less than 20\,$\mu$m per cycle. For a drop
of 3\,mm length this implies that it needs about 150 cycles to move by
its own length. The situation is very similar in Ref.~\cite{NKC09}
where for the parameters depicted in their Figs.~1 and 2 it takes
about 50 vibration cycles to move the drop by its own length. Note,
that their Fig.~1 does not show snapshots taken during a single cycle,
but selected pictures from 70 cycles.

In terms of the number of cycles needed to move the drop by its own
length our results are very close to the experimental findings: At a
period of $T=100$ we have a scaled velocity of about
$0.1\times10^{-3}$ (our Fig.~\ref{fig:vT}(a)); for $a_0=10$ and
$\beta=0.2$ this gives $\langle v\rangle=2\times10^{-3}$ and an
advancement per cycle of $\Delta x=0.2$. For a drop like the one in
our Fig.~\ref{fig:evol1} this implies that it takes about 190 cycles
to move it by its own length. Employing instead the accelerations of
15$g$ as in \cite{BED07} it takes about 90 cycles. This is a fair
agreement between experiment and lubrication theory.
One may as well discuss the time scale $t_0$ employed in the
non-dimensionalisation and use it to compare the relevant vibration
frequencies in the model and in the experiment.  

%

Finally, we would like to discuss the relation of the discussed
ratchet mechanisms to transport a continuous phase to 'particle
ratchets'.  We do this by pointing out some formal analogies between
the employed thin film equation and a Fokker-Planck equation for
transport of discrete objects in a particle ratchet. 

The film thickness evolution equation 
\begin{equation}
\partial_t\,h\,=\,-\partial_x\,\left\{Q(h)\partial_x\left[\partial_{xx}h + P(h,x,t)\right]
+ Q(h) F(t)\right\}.
\mylab{eq:film2}
\end{equation}
shows similarities to a Fokker-Planck equation for interacting
particles in an external ratchet potential \cite{SMN04,SMN05}.
Its non-dimensional form is
\begin{equation}
\partial_t\,W\,=\,-\partial_x\,\left\{ W\partial_x P(W,x,t) + W F(t) \right\}
\mylab{eq:fok}
\end{equation}
with
\begin{equation}
P(W,x,t)=-T(t)\log W - \tilde{g} W - U(x,t)
\mylab{eq:fok2}
\end{equation}
In the particular case, it is an evolution equation for the
one-particle distribution function $W(x,t)$, where $U(x,t)$ is a
flashing ratchet potential (note that the one used in
\cite{SMN04,SMN05} is static) 'rocked' periodically by $F(t)$. The
function $T(t)$ stands for a periodic modulation of the
non-dimensional temperature, and $\tilde{g}$ is a non-dimensional
interaction parameter. One notes that both, the first order drift or
transport term and the second order diffusion term are modulated in
time. Note that $U(x,t)$ could as well be incorporated into the drift
term instead of the diffusion term.

However, differing from \cite{SMN04,SMN05} our cases (ii) and (iii) do
not incorporate an external potential.  The absence of a spatial
ratchet potential is compensated by strongly nonlinear prefactors of
the diffusion and the drift term that result in mean flow even for
harmonic (i.e., time-symmetric driving).  In the context of 'particle
ratchets' related systems are studied \cite{SMN04b,Cole06} where a
mean flux is created without spatial ratchet potential but through the
response of two different (weakly) coupled degrees of freedom to
time-periodic driving. The degrees of freedom are related to two
different interacting particle species in \cite{SMN04b}, and to
pancake vortices and Josephson vortices in layered superconductors in
\cite{Cole06}.  However, in contrast to our case in
\cite{SMN04b,Cole06} the driving is time-asymmetric.  Time-symmetric
driving can as well induce mean transport in a 'particle ratchet'
without spatially varying potential if the coupling of the different
degrees of freedom is sufficiently nonlinear. The only example we are
aware of is the ``vortex diode'' discussed in \cite{SaNo02}.
Note finally, that similar 4th order terms as the capillarity term in
Eq.~(\ref{eq:film2}) may be included in higher order Fokker-Planck
equations.
%

\acknowledgments

This work was supported by the European Union under grant
PITN-GA-2008-214919 (MULTIFLOW).


%
%

%
\end{document}